\documentclass[prb,fleqn,showpacs, twocolumn, prb, superscriptaddress]{revtex4}
\usepackage{epsfig}
\usepackage{graphicx}
\usepackage{color} 
\usepackage{amsfonts}

\newcommand{\ct}{\cite}
\newcommand{\bi}{\bibitem}
\newcommand{\be}{\begin{equation}}
\newcommand{\ee}{\end{equation}}
\newcommand{\ba}{\begin{eqnarray}}
\newcommand{\ea}{\end{eqnarray}}
\newcommand{\al}{\alpha}

\newcommand{\non}{\nonumber}

\newcommand{\ket}[1]{|#1\rangle}
\newcommand{\de}{\delta}
\newcommand{\la}{\lambda}

\begin{document}
\title{Loschmidt echo with a non-equilibrium initial state:  early time scaling and enhanced decoherence}
\author{Victor Mukherjee}
\email{victor.mukherjee@cea.fr}
\affiliation{Institut de Physique Th\'{e}orique, CEA Saclay, F-91191 Gif-sur-Yvette Cedex, France}
\author{Shraddha Sharma}
\email{shrdha@iitk.ac.in}
\affiliation{Department of Physics, Indian Institute of Technology Kanpur, 
Kanpur 208 016, India}
\author{Amit Dutta}
\email{dutta@iitk.ac.in}
\affiliation{Department of Physics, Indian Institute of Technology Kanpur,
Kanpur 208 016, India}

\begin{abstract}
 
We study the Loschmidt echo (LE)  in a central spin model in which a central spin is globally coupled to an environment (E) 
which is subjected to a small and sudden quench at $t=0$ so that {its
state  at $t=0^+$, remains the same as the ground state of the initial environmental Hamiltonian before the quench; this leads to a non-equilibrium
 situation}. This state now evolves with two Hamiltonians, the final Hamiltonian following the
quench and its  modified version which incorporates  an additional term  arising due to the coupling of the central spin to the environment. Using  a generic 
short-time scaling of the decay rate,
we establish that in the early time limit, the  rate 
of 
decay of the LE  (or the overlap
between two states generated from the initial state evolving through two channels ) close to the quantum critical 
point (QCP) of E  is independent of the quenching.  We do also study the temporal evolution of 
the LE and establish the presence of a crossover to a situation where the quenching becomes irrelevant. 
In the limit of large quench amplitude the
non-equilibrium initial condition  is found to result in a  drastic increase in decoherence at large times, even far away from a QCP. 
These generic results are verified analytically as well as numerically, choosing E to be a transverse Ising chain where the 
transverse field is suddenly quenched. 

\end{abstract}
\pacs{03.65.Yz, 05.50.+q, 05.70.Jk, 64.70.qj, 64.70.Tg, 75.10.Jm}
\maketitle 

The emergence of the classical world from the quantum world, namely decoherence, or the quantum-classical transition 
through reduction of a pure state to a mixed state has been a subject of perpetual interest to
the physics community \ct{zurek91,haroche98,zurek03,joos03}. The concept of the LE has been proposed in connection to this quantum-classical transition in
 quantum chaos to describe the hypersensitivity of the time evolution of
a system to the perturbation experienced by its surrounding
 \ct{peres84,peres95,zurek94,jalabert01,karkuszewski02,cucchietti03, diez10, venuti11}. The LE is defined as follows: 
if a quantum state {$|\psi \rangle$ evolves with two Hamiltonians
$H$ and $H'$, respectively, the LE is the measure of the overlap given by
$$ {\cal L} (t) = |\langle \psi|\exp(iH't) \exp(-iH t)|\psi \rangle|^2. $$}

In recent years, the temporal evolution of the LE has been studied in the vicinity of a QCP. In this context, the central spin model (CSM) where a central spin (CS) is coupled globally to
all the spins of  an environment (E) which is chosen to be a transverse Ising
chain  has been introduced \ct{quan06}. The CS is assumed  to be in a pure state initially while the spin chain is in the ground state. The interaction between
the central spin and the environment effectively leads to two Hamiltonians which provide  
two channels of time evolution of  the environmental ground state and lead to a decay in the LE. 
It has been reported  that in the limit of weak coupling between the central spin and the environment, the LE shows a sharp decay close to the QCP of the spin chain
and right at the QCP, it shows a collapse and revival as a function of time $t$. This collapse
and revival of the LE can be taken to be an indicator of the proximity to a QCP. It can also be shown that { the CS makes} a transition to a mixed state when the LE vanishes. 
 
 The CSM has been generalized
 to a more generalized environmental Hamiltonian \ct{yuan07}, to the limit of strong coupling \ct{cucchietti07}, 
{to the case when the interaction between the CS and 
E is local
(i.e., CS coupled to a single spin of the E)
rather than global \ct{rossini07} and also in the realization of a Schr\"odinger magnet\ct{rams12}}. In  the short-time limit, one finds  a sharp decay of the LE 
given by the Gaussian form{\ct{peres84}} ${\cal L}(t) \sim \exp(-\alpha t^2)$,  where the decay
rate $\alpha$ is expected to capture the universality associated with the QCP of the E \ct{rossini07, zhang09,venuti10,sharma12}.
In a recent experimental
study {with} NMR quantum simulator, it has been observed that for a fixed short time the LE approaches a minima
in the vicinity of the QCP of an antiferromagnetic Ising chain \ct{zhang09}. The connection between the dynamic LE approach
with the static fidelity approach has also been established \ct{zanardi06}.

In this paper, we study the temporal evolution of the LE following a sudden quench of the E  
by a perturbation at $t=0$ so that the {state of E} at $t=0^{+}$, 
is an eigenstate of the initial Hamiltonian
($H_I$) but not of the final Hamiltonian ($H_F$). Our aim here is to study how a sudden quench 
of the E influence  the LE (or the decoherence
of the CS) both in the early time limit and also in its time evolution. 
Denoting the excited (ground) state 
of CS by $|e\rangle$ ($|g\rangle$) one can write down 
generic Hamiltonian of the system and environment in the form \ct{quan06}
\ba
H(\lambda + \de,g ) = H_0 + \left(\lambda + \de |e\rangle \langle e| \right)V_{\la} + g V_g,
\label{hamil1}
\ea
Here, $H_0 + \la V_{\la}$ is the initial Hamiltonian $H_I(\la)$, of the E where $\la$ is the measure of the deviation from
the QCP ($\la=0$) while  $g V_g$ denote the perturbing term responsible for the sudden quench so
that the final Hamiltonian $H_F(\la,g) = H_I(\la) + g V_g$; $V_\la$ and $V_g$ do not necessarily  commute  with each
other. {Unless otherwise stated $g$ is always taken to be positive.} The term $\de (|e\rangle \langle e|) V_{\la}$ is the global coupling between the CS and the E where we have assumed that the CS couples with the E only 
when it is in the excited
state $|e\rangle$ and also the small $\de$ limit throughout. Initially  ($t = 0$) the CS  is assumed to be in a pure  
state, 
$ |\phi_S\rangle = c_g|g\rangle + c_e |e\rangle, $ with the $|c_g|^2 + |c_e|^2 =1$. On the other
hand, since the E undergoes a sudden quench at $t=0$, it is in the ground state of the Hamiltonian
$H_I(\la)$ at $t=0^+$ denoted by $|G(\la,g=0)\rangle$.

 As a consequence of the coupling between the environment and the CS, the state  $|G(\la,g=0)\rangle $ starts evolving following two separate hamiltonians
 $ H_F(\la + \de,g) = H_0 + (\lambda + \de) V_{\la} + gV_g $ and $ H_F(\la,g)= H_0 + \lambda V_{\la} + gV_g $ depending on
  the state of the CS.  At an arbitrary time $t$, we can 
therefore write the state of the composite system as
\ba
|\psi(t)\rangle = c_g|g\rangle \otimes |\phi_g(t)\rangle  + c_e |e\rangle \otimes |\phi_e(t)\rangle,
\label{psit}
\ea
where $|\phi_g(t)\rangle = \exp(-iH_F(\la,g) t )|G(\la, g=0)\rangle$ and  $|\phi_e(t)\rangle = \exp(-iH_F(\la + \de,g) t )| G(\la, g=0)\rangle$.
 The LE at time $t$ is defined as
\ba
&&{\cal L}_q(\la,g, t) = |\langle \phi_g(t)|\phi_e (t)\rangle|^2 \non\\
&=& |\langle G(\la,g=0)| e^{iH_F(\la,g)t}e^{-iH_F(\la+\de,g)t}|G(\la,g=0)\rangle |^2.
\label{eqlegen}
\ea
To contrast with the equilibrium situation ($g=0$) \ct{quan06}, we note that in the present  case $ |G(\la,g=0)\rangle$ is not an eigenstate of $H(\la,g)$. Throughout we shall denote the  LE for the equilibrium case (no quenching)
as ${\cal L}$.

 Having introduced our model, let us briefly summarize our results. In the following, we propose
 a general scaling for decay rate $\al$ of the LE in the short time limit close to the QCP and
 show that the scaling is unaffected by  perturbation $V_g$. We however show that
 there is a crossover to the equilibrium behavior for a limit of the quenching amplitude $g$ when
 the correction in the LE due to the quenching becomes irrelevant. When the quenching is
 relevant, it deeply influences the temporal evolution of the LE and shows a faster decay
 as a function of time  even when the E is quenched to the QCP and one observes a conspicuous
 absence of the proper  revival when the environment is chosen to be a transverse Ising chain.

 To arrive at the scaling relation valid in the short time limit, we assume the limit $\de \to 0$
 and truncate the exponentials in Eq. (\ref{eqlegen}) up to the order $t^2$. One can then
 express the LE in the form  $
{\cal L}_q(\la,g, t) \approx 1 - \al t^2 \approx e^{-\al t^2},$
where
$\al = \delta^2 \left[\langle V_{\la}^2 \rangle - \langle V_{\la} \rangle^2 \right],
$ where $\langle ...\rangle$ implies the expectation value in the {state} $|G(\la,g=0) \rangle$ \ct{peres84}.
 This implies that the small time limit 
behavior of the echo is independent of the quench amplitude $g$.

Let us  note that the operator  $\la V_{\la}$ is a relevant or marginal perturbation that drives the E away from the
gapless QCP thereby generating a gap in the spectrum \ct{grandi10}; it is then expected to  scale in the same way as energy, 
implying $\la V_{\la} \sim \la^{\nu z}$ for
 $\la \gg L^{-1/\nu}$,
 where $\nu$ and $z$ are the 
correlation length and dynamical exponents, respectively, characterizing the QPT in the E	 driven 
by $\la$ . We  eventually find  the general scaling form
\ba
\al \sim \de^2  \la^{2\nu z - 2 } ~ (\la \gg L^{-1/\nu} );~\al \sim\de^2  L^{2/\nu - 2z}~(\la \ll L^{-1/\nu} ),
\label{L2}
\ea
where $L$ is the linear dimension of the system. However, we note that the above scalings are valid only as long as $2/\nu - 2z > d$,
 where $d$ is the spatial dimension. Otherwise the contribution from the low 
energy modes become sub-leading  and  that from the higher energy modes dominate resulting
in the scaling $\al ~\sim L^d$. In deriving the above relation, we have assumed that the CS
 is coupled to the  E through the operator $V_{\delta}$ which is chosen to be equal to $V_{\la}$
 that drives the system away from the QCP. Otherwise, the critical exponent $\nu_{\delta}$, that
 determines the scaling of correlation length when the system is moved away from the QCP ($\la=0$)
 due to the perturbation $V_{\delta}$ (correlation length, $\xi \sim \de^{-\nu_\de}$), will replace $\nu$ in Eq.~{\ref{L2}}. 
{ We note that studies for the $g = 0$ case in the recent past have yielded similar results\ct{venuti10}.}

Let us now reconcile the scaling given in Eq.~(\ref{L2}) with that presented in reference [\onlinecite{quan06}]
when $\nu=z=1$.
In the short time limit, if one considers only low energy modes up to a cut off momentum $\hat k$
(assuming that the critical mode $k_c=0$), we find that the energy gap scales $\hat k ^z \sim L^{-z}$
close to the QCP. With $\la V_{\la} \sim L^{-z}$, we find identical result $\al \sim \la^{-2} L^{-2z}$
to that given in \ct{quan06} which has also been verified for situations with $z \neq 1$ \ct{sharma12}.
Scaling relations presented in 
(Eq. \ref{L2}) are  also in congruence with the dimensional analysis noting that $\de \sim \la \sim L^{-1/\nu}$ and $t \sim \la^{-\nu z} \sim L^{z}$.

\begin{figure}[ht]
\begin{center}
\includegraphics[width=7.7cm]{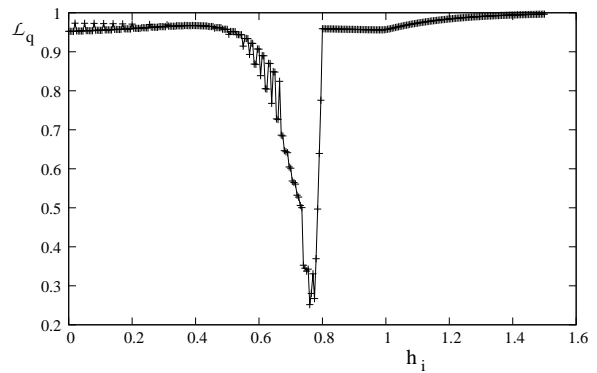}
\end{center}
\caption{Variation of ${\cal L}_q(t)$ as a function of $h_i$ for {$L = 200$, $\delta = 0.025$, $g = 0.2$ and $t=1$, as obtained numerically
 from Eq. (\ref{L_analytical})}. There is a sharp dip around the QCP.}
\label{Fig1}
\end{figure}

\begin{figure}[ht]
\begin{center}
\includegraphics[width=7.7cm]{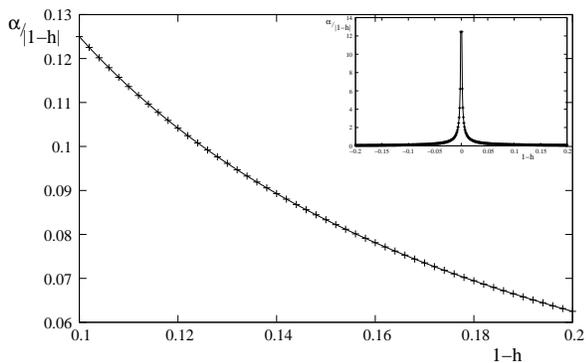}
\end{center}
\caption{Variation of $\alpha/(1 - h)$  as a function of $\left(1 - h\right)$ inside the ferromagnetic region for 
$h_i = h_f = h$, $g = 0$, $L = 500$, $t = 0.005$ and $\delta = 0.005$. $\alpha/(1 - h)$ falls as $(1 - h)^{-1}$ implies $\alpha$ is independent of $(1 - h)$.
{ Inset shows $\al/|1 - h|$ varies symmetrically on either side of the QCP.}}
\label{Fig2}
\end{figure}

\begin{figure}[ht]
\begin{center}
\includegraphics[width=7.7cm]{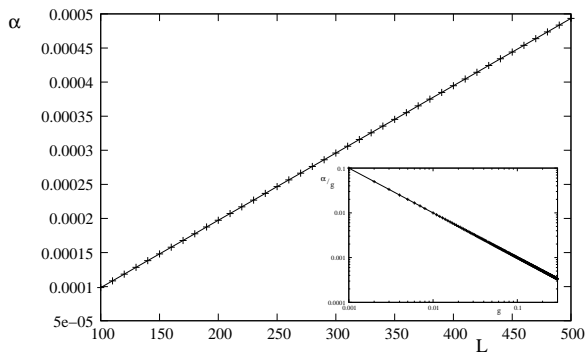}
\end{center}
\caption{The decay rate  $\alpha$ as a function of system size $L$ for $g = 0$, $h_i = h_f = 0.999$, $t = 0.005$, $\delta = 0.0001$ is shown. {The contribution from the 
higher energy modes makes $\al$ increase linearly with $L$. Inset shows the variation of $\alpha/g$ as a function of $g$ for $L = 100$, $t = 0.0001$, $\delta = 0.001$ $h_i = 0.8$, $h_f = h_i + g$. $\alpha/g$
 falls as $g^{-1}$ implying that $\alpha$ is independent of $g$.}}
\label{Fig3}
\end{figure}

Let us shift our attention  beyond the short-time limit to explore the effect of quenching on the LE at an arbitrary time but for small values
of $g$.  
Assuming that
$|(\partial G (\la,g) \rangle /\partial g)$ exists everywhere 
{and using   Eq.~(\ref{eqlegen}), one finds 
 \ba
{\cal L}_q(\la,g,t) \approx {\cal L}(\la,0, t) + g\frac{\partial {\cal L}_q(\la,g,t)}{\partial g}|_{g = 0} ~~...,
\label{eq_le_time}
\ea
where ${\cal L }(\la, 0,t)$ is the Loschmidt echo in the equilibrium situation $g = 0$ when the environment is 
initially in the state $\ket{G(\la, 0)}$ which is the ground state of $H_F(\la,g = 0) = H_I(\la)$. Hence,
$\Delta_{{\cal L}} = {\cal L}_q(\la,g, t) - {\cal L}(\la, 0,t)$ varies linearly in $g$ if  $\partial {\cal L}_q(\la,g,t)/\partial g|_{g = 0}$ 
exists and is non-zero. Demanding the Loschmidt echo to be dimensionless, we  find  that $\partial {\cal L}_q(\la,g,t)/\partial g|_{g = 0}$ must  have the dimension 
$g^{-1} \sim L^{1/\nu_g}$ where $\nu_g$ is the correlation length exponent corresponding to the operator  $V_g$.
 We therefore propose a very interesting crossover scenario;
in case the quenching parameter $g \gg L^{-1/\nu_g}$, the quenching influences the LE as given in Eq.~(\ref{eq_le_time}). In 
the other limit 
$g \ll L^{-1/\nu_g}$, the correction due to the quenching becomes irrelevant and one recovers the results for the equilibrium 
case \ct{quan06}.
{ Eq. (\ref{eq_le_time}) also shows that $\Delta_{{\cal L}}$ depends on the sign of $g$, i.e., 
the difference  in the LE (compared to the $g=0$ case)  depends non-trivially on the direction of
 quench for small values of $|g|$.}{ However, in the limit of large $|g|$ ($|g| \gg L^{-1/\nu_g}$), ${\cal L}(\la,0, t)$  is in general finite 
far away from the QCP; this can   be easily seen by
 expanding ${\cal L}(\la,0, t)$ for small $\delta$ and noting that $\partial \ket {G(\la,0)}/\partial \la$ is large only near the QCP 
and small otherwise\ct{quan06, rossini07}. }

On the other hand, in order to investigate the dependence of ${\cal L}_q(\la,g,t)$ on $g$, 
 we expand $\ket{G(\la,0)}$ in Eq. (\ref{eqlegen}) as 
$\ket{G(\la,0)} = \sum_{n} a_n \ket{\Xi_n(\la,g)}$ in terms of the eigenstates $\ket{\Xi_n(\la,g)}$  of $H_F(\la,g)$ (with corresponding probability 
amplitudes $a_n$ and eigenenergies 
$\epsilon_n(\la,g)$). Increase in $g$ increases the
number of oscillatory terms contributing to the expansion of ${\rm{exp}}\left(iH_F(\la,g)t\right)\ket{G(\la,0)} =
 \sum_{n}{\rm{exp}}\left(i\epsilon_n(\la,g)t\right) a_n \ket{\Xi_n(\la,g)}$, which finally leads to decrease in value of ${\cal L}_q(\la,g,t)$ at any finite 
time. Therefore in the limit of large $g$ we can
 expect $\Delta_{{\cal L}}$ to be negative at large times irrespective of sign of $g$, or in other words, a sudden quench of an 
environment coupled to a qubit leads to enhancement in decoherence as compared to the case with equilibrium initial condition. 
We note that the dependence of ${\cal L}_q(\la,g,t)$ and $\Delta_{{\cal L}}$ on $g$ as discussed above are not necessarily related to a QCP, and should be  prominent even  away from it. This is one of the central results of our work and it shows that equilibrium 
initial condition is absolutely necessary for a coherent time evolution in an open quantum system, even far away from the QCP. It is worthwhile to note that recent studies have shown that non-equilibrium critical dynamics (achieved by a slow variation of a parameter) of the environment across its QCP also leads to 
dramatic enhancement in decoherence\ct{damski11}.

We elucidate the above generic theories using the exactly solvable example of a spin $1/2$ in an environment of transverse Ising spin chain 
so that  the composite Hamiltonian is
\ba
H_I{(h+\de)} = -\sum_{j=1}^L \left(\sigma_j^x \sigma_{j+1}^x + h\sigma^z_j + \delta|e\rangle\langle e|\sigma^z_j\right).
\label{hamilIsing}
\ea
The environmental spin chain undergoes a QPT at $h=h_c= \pm 1$ \ct{lieb61,dutta10}. At time $t = 0$,
the transverse field is suddenly changed from the initial value $h_i$ to the final value $h_f=h_i+g$ so that
the initial state  is the  ground state $|G(h_i,0)\rangle$ of the Hamiltonian $H_I(h_i)$. We are
 considering the case when the quenching along direction of  the transverse field that
drives the quantum transition (i.e., $V_g=V_{\la}$).

The Loschmidt echo at time $t > 0$ can be calculated exactly in the form
\ba
&~&{\cal L}_q(t)  
= |\langle G(h_i)|e^{iH(h_f)t}e^{-iH(h_f + \delta)t}|G(h_i)\rangle|^2 = \prod_{k>0}{F_k}\non \\
&~&= \prod_{k>0}|A_ke^{i\Delta_k t} + B_ke^{-iS_kt} - C_ke^{iS_k t} + D_ke^{-i\Delta_k t}|^2,
\label{L_analytical}
\ea 
where \\$A_k = \cos{\alpha_k(h_{i},h_f)} \cos{\alpha_k(h_{i},h_f + \delta)}\cos{\alpha_k(h_f,h_f+\delta)}$ \\ 
$B_k = \cos{\alpha_k(h_{i},h_f)} \sin{\alpha_k(h_i,h_f + \delta)}\sin{\alpha_k(h_f,h_f+\delta)}$, \\ 
$C_k = \sin{\alpha_k(h_{i},h_f)} \cos{\alpha_k(h_{i},h_f + \delta)}\sin{\alpha_k(h_f,h_f+\delta)}$, \\
$D_k = \sin{\alpha_k(h_{i},h_f)} \sin{\alpha_k(h_{i},h_f + \delta)}\cos{\alpha_k(h_{f},h_f + \delta)}$, \\ 
$\Delta_k = \varepsilon_k(h_f + \delta) - \varepsilon_k(h_f)$ and $S_k = \varepsilon_k(h_f + \delta) + \varepsilon_k(h_f)$, \\
$\alpha_k(m,n) = (\theta_m - \theta_n)/2$, $\theta_m = \tan^{-1}{\sin(k)/(m - \cos k)}$ and
$\varepsilon_k(h) = 2\sqrt{\left(h - \cos{k}\right)^2 + \sin^2 k}$. We note that the limit $g = 0$ ($i.e., h_i = h_f$),
we recover the result of the reference [\onlinecite{quan06}], where we have used periodic boundary conditions such that the decoupled momentum modes $k$ are 
quantized as\ct{lieb61} $k = 2\pi p/L$, $p = 0, 1, 2,...L$.

We shall now analyse  Eq. (\ref{L_analytical}) numerically to study the proximity to the QCP.   
Fig.~(1) shows that ${\cal L}_q(t)$ plotted as a function of $h_i$ shows a minimum near $h_i = h_c-g=1 - g$, 
thus detecting the presence 
of a quantum critical point  even when the environment is suddenly quenched.

Let us now verify whether the scaling relations of $\al$ in the early time limit given above holds true 
in this specific case. We present results in {Fig. (\ref{Fig2}); the variation of $\al$ 
is plotted as a function of $\la=1-h$ close to the QCP with $g=0$};
we find the results are in perfect agreement with the proposed scaling. In inset of Fig.~(2), we
show that the short-time decay  also depends on the symmetry of the phase on either side of the QCP \ct{zhang09}. Moreover, $\al$
is linear in $L$ which is essentially a contribution from the high-energy modes as we show
in Fig. ~(\ref{Fig3}) where we also show that the decay rate $\al$  does not depend on $g$
for $g \neq 0$.

In Fig. (\ref{Fig5}), we plot the behavior of ${\cal L}_q(t)$ as a function of time, as obtained numerically using Eq. (\ref{L_analytical}). Interestingly, 
for nonzero $g$, ${\cal L}_q$ initially decays  for small time before rising again and showing small oscillations at large times. On the other hand the echo stays 
almost close to unity away from the QCP for $g = 0$. One therefore finds that quenching deeply
influences the collapse and revival of the LE following a quench. In fact, absence of perfect revival
(even when the E is quenched to the QCP with $h_f+ \de=1$) and very fast decay of the peak in
${\cal L}$ establishes this conjecture. Moreover,
 we consider the limit $\de \ll g$  when
$\sin{\alpha_k(h_f,h_f+\delta)} \ll \cos{\alpha_k(h_f,h_f+\delta)}$ and one gets
\ba
F_k &\approx& |A_ke^{i\Delta_k t} + D_ke^{-i\Delta_kt}|^2.
\label{echo_time}
\ea
Near the critical momentum mode $k = 0$, 
$\Delta_k = \varepsilon_k(h_f + \delta) - \varepsilon_k(h_f) \sim 2(h_f + \delta - h_f) = 2\delta$ for $g \gg L^{-1}$, eventually leading to 
\ba
F_k \approx (A_k + D_k)^2\cos^2(\Delta_k t) + (A_k - D_k)^2\sin^2(\Delta_k t).
\label{t0del}
\ea
The time at which the maxima occurs can be determined by the condition $\partial F_k/\partial t = 0$ 
which yields $t_{n,k} = n\pi/\Delta_k \sim \de^{-1}$ ($n = 1,2, \cdots$).
However, the momentum dependent $t_{n,k}$ results in the irregular oscillatory temporal behavior of ${\cal L}_{q}(t)$, as shown in Fig. (\ref{Fig5}). 
In Fig.~(\ref{Fig6}), we show the time $t_0$ at which the first
maxima occurs indeed scale with $\de^{-1}$.

\begin{figure}[ht]
\begin{center}
\includegraphics[width=7.7cm]{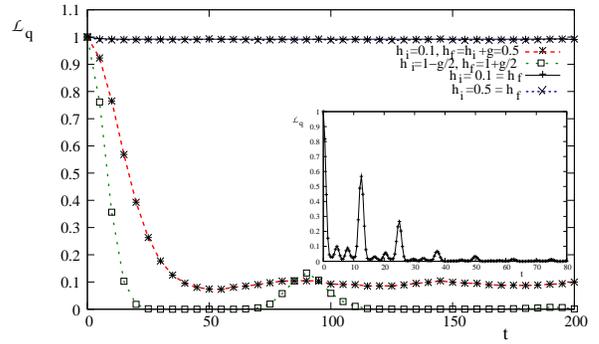}
\end{center}
\caption{This figure shows the variation of ${\cal L}_q(t)$
with time for {$L = 100$, $\delta = 0.025$ and various values
of $g$, $h_i$ and $h_f$
(as marked in the figure). The blue and black curve corresponds to $g=0$
whereas the red and green curve marks values corresponding to $g=0.4$. The green curve corresponds to crossing the QCP
 Inset: ${\cal L}_q(t)$ with time for $L=100$, $\delta=0.025$, $h_i=0.575$ and $g=0.4$,
so that  $h_f+\delta=1$.}}
\label{Fig5}
\end{figure}

\begin{figure}[ht]
\begin{center}
\includegraphics[width=7.7cm,angle=0]{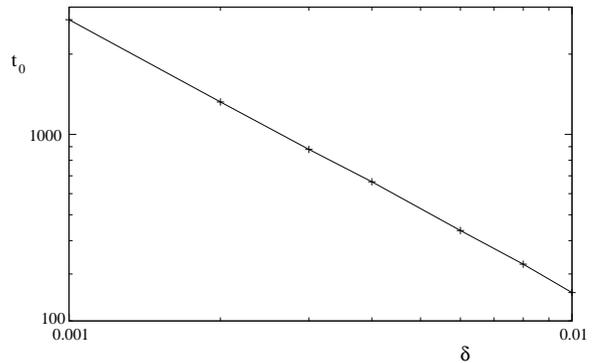}
\end{center}
\caption{Time $t_0$ plotted as a function of $\delta$ for $L = 100$, $h_i = 0.95$, $h_f = 1.05$. 
As expected, $t_0$ falls linearly with $\de$. }
\label{Fig6}
\end{figure}

Let us now estimate  characteristic value of $g = g_0 \sim L^{-1/\nu_g}$; for $g >g_0$, we expect to see the influence of the non-equilibrium initial 
condition.  Inspecting,  Eq.~(\ref{L_analytical}), we expect  a cross-over when the contributions from the terms $C_k$ and $D_k$  become negligible as 
compared  to those coming from $A_k$ and $B_k$. This is ensured when $\sin{\alpha_k(h_{i},h_f)} \approx \cos{\alpha_k(h_{i},h_f)}$. Using the initial 
and final parameters to be  
$h_i = 1 - g/2$, $h_f = 1 + g/2$,  we {find  $g_0 \sim L^{-1}$, as is expected from our previous analysis because $\nu_g = 1$,
 in the present case.}
\begin{figure}[ht]
\begin{center}
\includegraphics[width=7.7cm]{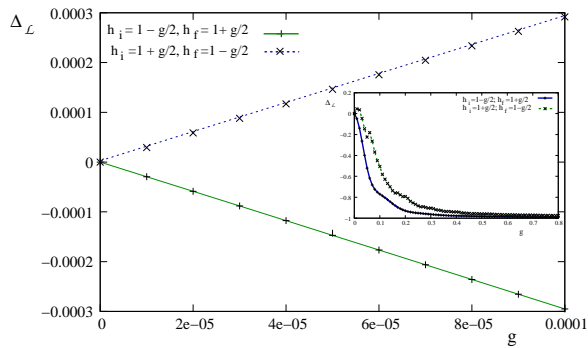}
\end{center}
\caption{{ $\Delta_{\cal L}$ decreases (increases) 
linearly with $g$ 
for $L = 200$, $t = 20$, $\delta = 0.025$, $h_i = 1 - g/2$ and
$h_f = 1 + g/2$, ($h_i = 1 + g/2$ and
$h_f = 1 - g/2$) in the limit of small $g$. Inset : $\Delta_{\cal L}$ asymptotically
decreases (to $\approx -1$) for large $g$, with $L = 200$, $t = 20$, $\delta
= 0.025$, $h_i = 1 \mp g/2$ and
$h_f = 1 \pm g/2$, confirming that for large $g$, the coherence disappears completely irrespective
of proximity of a QCP and direction of quenching. }}
\label{Fig7}
\end{figure}
 In the limit of small $g$
it is straightforward to verify numerically that the difference $\Delta_{{\cal L}}$ scales linearly with $g$ as proposed in the generic case
 (see Fig. (\ref{Fig7})). {Interestingly, as expected from our earlier discussions, even though the sign of 
$\partial \Delta_{{\cal L}} /\partial g$ is reversed for small $|g|$ when direction of quench is reversed, in contrast in 
 the limit of large $|g|$ ($|g| \gg g_0$) $\Delta_{{\cal L}}$ 
asymptotically decreases to $-1$ irrespective of the direction of quenching, signifying almost complete loss of coherence 
(see Fig. (\ref{Fig7})) .}

In conclusion, we have studied the LE both in the early time limit and as a function of time following
a sudden quench of the environment. We have proposed a generic early time scaling
which is independent of the quenching and determined entirely by the scaling dimension of the operator $V_{\de}$  that couples the 
CS to the E and the dynamical exponent $z$ which is
a characteristic of the associated QCP of the environment. Moreover, our study predicts an interesting crossover
behavior. {We also show that non-equilibrium initial condition can lead to drastic increase in decoherence in open 
quantum systems, even away from a QCP.} 
For the specific example in which a transverse Ising chain is chosen as the E, we 
not only verify the generic scaling relations but also establish that the perfect revival of LE does not occur. However, in this  
example we have chosen $V_{\la}=V_{g}=V_{\de}$. It would be instructive to verify when the CS is
coupled to the E through a longitudinal field term ($\de |e \rangle \langle e| \sum_i \sigma_i^x$) or
the E is quenched by a term $g  \sum_i \sigma_i^x$. Although the model then becomes non-integrable, the exponents $\nu_g$ and
$\nu_{\de}$ will be different from $\nu=1$ (in fact, the  critical exponent $\nu$  when the QCP is perturbed by a longitudinal field  
 is\ct{pollmann10} $8/15$ ) and are expected to  appear in the early time scaling and the crossover relation. Further, this
work can also lead to additional studies on effects of different quenching schemes on decoherence and generalization
 to finite temperatures may also yield interesting results. Finally, recent experiments on Loschmidt echo using NMR quantum simulator
\ct{zhang09} and also other notable experiments on non-equilibrium quantum dynamics (for example see Ref. [\onlinecite{sadler06}]) have created the possibility of experimental verification 
of our results in near future.

AD acknowledges CSIR, New Delhi, for support and SS for Junior Research Fellowship.

\end{document}